\documentclass{article}

\usepackage[english]{babel}

\usepackage[letterpaper,top=2cm,bottom=2cm,left=3cm,right=3cm,marginparwidth=1.75cm]{geometry}

\usepackage{amsmath}
\usepackage{graphicx}
\usepackage[colorlinks=true, allcolors=blue]{hyperref}
\usepackage[numbers]{natbib}
\usepackage{tikz}
\usepackage{caption}
\usepackage{capt-of}
\usepackage{subcaption}
\usepackage{float}
\usepackage{array}
\usepackage{hyperref}
\usepackage{tabularx}
\usepackage{xcolor}
\usepackage{hhline}
\usepackage{colortbl}
\usetikzlibrary{positioning,shapes.geometric,fit,arrows,shapes.misc,patterns,arrows,decorations.pathreplacing,matrix}
\usepackage{xurl}
\usepackage{mathtools}
\usepackage{relsize}
\usepackage{array,ragged2e}
\newcolumntype{P}[1]{>{\RaggedRight\arraybackslash}p{#1}}
\newcommand*\samethanks[1][\value{footnote}]{\footnotemark[#1]}

\title{Supernodes}
\author{S.~Y. Chan\thanks{Deakin University, Geelong, Australia, School of Information Technology, Faculty of Science Engineering \& Built Environment, \underline{Australia}}        
    \and    
    K. Morgan\samethanks
    \and
    N. Parsons\thanks{Deakin University, Geelong, Australia, School of Psychology, Cognitive Neuroscience Unit, \underline{Australia}}
    \and
    J.Ugon\samethanks[1]}
\date{}

\begin{document}
\maketitle

\begin{abstract}
In this paper, we present two new concepts related to subgraph counting where the focus is not on the number of subgraphs that are isomorphic to some fixed graph $H$, but on the frequency with which a vertex or an edge belongs to such subgraphs. In particular, we are interested in the case where $H$ is a complete graph. These new concepts are termed \emph{vertex participation} and \emph{edge participation} respectively.  We combine these concepts with that of the \emph{rich-club} to identify what we call a \emph{Super rich-club} and \emph{rich edge-club}. We show that the concept of vertex participation is a generalisation of the rich-club. We present experimental results on randomised Erd\"{o}s R\'{e}nyi and Watts-Strogatz small-world networks. We further demonstrate both concepts on a complex brain network and compare our results to the rich-club of the brain.  
\end{abstract}

\section{Introduction}
\label{sec:intro}

Networks are ubiquitous in nature, and are important in numerous scientific disciplines including social science, computer science and biology~\cite{costa2011,vandijk2012,goldenberg2010}. These networks arise from interactions between entities~\cite{newman2018} and can be represented as a graph, where vertices (nodes) model entities and edges model relationships between these entities. Graph metrics are used to analyse, measure, model and interpret these graphs. 

Due to the emergence of applications such as identifying network motifs and understanding biological networks~\cite{komusiewicz2018,maxwell2014,milo2002}, and more recently in social networks and brain \textsc{mri} imaging, there is growing interest in studying the substructures of networks by means of counting, enumerating, classifying, and analysing subgraph \emph{within} a graph~\cite{aparicio2014parallel,bera_et_al, chakaravarthy2016_counting, fomin2012, kloks2000, maugis2018,ribeiro2010parallel, ribeiro2019survey}.  

The subgraph counting problem~\cite{fomin2012, ribeiro2019survey} consists of finding the number of subgraphs of a large input graph $G$ that are isomorphic to a given pattern graph $H$. Subgraph counts can be used to describe distinctive features of a graph $G$, by identifying certain subgraphs (called \emph{motifs})~\cite{milo2002} that appear more frequently in $G$ than expected in a random graph. Counting subgraphs is a hard problem, since it generalises the subgraph isomorphism problem which is NP-complete~\cite{cook1971theorem}. In this paper, we propose new graph measures based on subgraph counting. These new measures are then applied to advanced magnetic resonance imaging (\textsc{mri}) data from the healthy human brain --- an example of a complex biological network.

\subsection{Brain Network and the Rich-Club.}
The brain can be modelled as a network where vertices represent regions and edge weights represent the amount of white matter connecting regions~\cite{sporns2010book}. In the study of the brain, graph theory has been applied to reveal important information about this complex network. For example,~\citet{bassett2006} showed the small-worldness of brain networks using graph metrics such as clustering coefficient and path length. Further,~\citet{heuvel2016comparative} applied graph metrics in comparing brain network topology across species. 

Studies~\cite{grayson2014,heuvel2011} have revealed some highly connected regions of the brain form large-scale hubs. These hubs are termed a \emph{rich-club}, and may reflect high-bandwidth information transportation between key brain areas. The \emph{density} of a graph $G$ is given by 

\begin{equation}
\delta(G) = \dfrac{|E(G)|}{\binom{|V(G)|}{2}}=\dfrac{2|E(G)|}{|V(G)|^2-|V(G)|}
\end{equation}
where $0\leq \delta(G) \leq 1$. The \emph{rich-club coefficient} $\phi(G,j)$ can be defined as the density of the subgraph $H$ induced by the vertices of degree greater than $j$ in the graph $G$. This coefficient measures to what extent these vertices form a rich-club. The concept of the rich-club was first introduced by~\citet{zhou2004} and has been applied on various networks~\cite{colizza2006,csigi2017,vaquero13} (i.e., social networks,internet networks, classroom networks). In recent years, there has been interest in elucidating these rich-clubs in the brain; as they may reveal information about the connectedness or indeed the importance, of specific brain regions ~\cite{bassett2009,gong2008,hagmann2008,heuvel2010,heuv2009}. This coefficient~\cite{colizza2006,mcauley2007,zhou2004} is given by:
\begin{equation}
\phi(G,j)=\frac{E_{>j}}{\binom{N_{>j}}{2}}
\label{eq:rccoef}
\end{equation}
where $E_{>j}$ is the number of edges  and $N_{>j}$ is the number of vertices in the induced subgraph $H$ respectively. There are multiple variations of the weighted rich-club coefficient for weighted graphs~\cite{alstott2014}. 

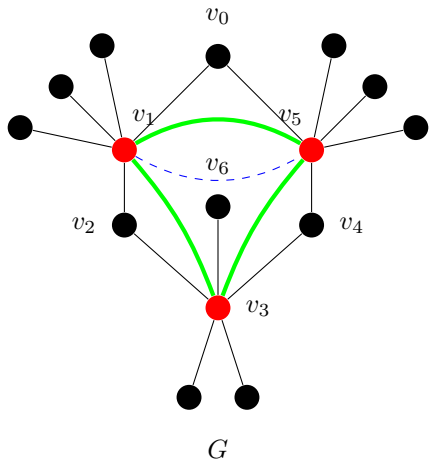
\begin{figure}
\centering
\begin{tikzpicture}
[scale=.35,auto=left,every node/.style={circle,fill=black},level distance=34mm, sibling distance=22mm]
\node[label=above:$v_{0}$] (n0) at (0,0) {}; 
\node[fill=none] (n1) [below of=n0] {};
\node[label=above:$v_{6}$] (n2) [below of=n1] {};
\node[fill=red, label=87:$v_{1}$] (n3) [below left=of n0] {} [grow=north west] child {node {}} child{node {}} child{node {}};
\node[fill=red,label=97:$v_{5}$] (n4) [below right=of n0] {} [grow=north east] child{node {}} child{node {}} child{node {}};
\node[fill=red,label=right:$v_{3}$] (n5) [below=of n2] {} [grow=south] child{node {}} child{node {}};
\node[label=left:$v_{2}$] (n6) [below of=n3] {};
\node[label=right:$v_{4}$] (n7) [below of=n4] {};
\node[fill=none, label=below:$G$] (n8) [below=of n5] {};
  
\draw (n0) -- (n3);
\draw (n0) -- (n4);
\draw (n3) -- (n6);
\draw (n5) -- (n7);
\draw (n5) -- (n6);
\draw (n2) -- (n5);
\draw (n4) -- (n7);
\draw[dashed, blue] (n3) to[bend right] (n4);
\draw[ultra thick, green] (n3) to[bend left] (n4);
\draw[ultra thick, green] (n5) to[bend left=10] (n4);
\draw[ultra thick, green] (n3) to[bend left=10] (n5);
  
\end{tikzpicture}
\caption{An illustration of a graph $G$ with a \emph{rich-club} marked in red for $j=4$. Let $G'$ be the graph $G$ with an additional dashed edge and $G''$ be the graph $G$ with the addition of the green edges. The rich-club coefficient for $G$, $G'$ and $G''$ are 0, $\frac{1}{3}$ and 1 respectively.}
\label{fig:example RC}
\end{figure}

The vertices belonging to the rich-club are selected based on the degree of the vertices, for degree $>j$. These vertices are known as the \emph{rich-club members}. In Figure~\ref{fig:example RC}, the members of a rich-club are highlighted in red. We illustrate different rich-club coefficients by the addition of the dashed edge and green edges respectively. Here $G'=(V,E')$  where $E'(G)=E(G)\cup\lbrace \{v_{1},v_{5}\}\rbrace$ and $G''=(V,E'')$ where $E''=E(G)\cup\{\{v_{1},v_{5}\},\{v_{1},v_{3}\},\{v_{3},v_{5}\}\}$. Thus, we have $\phi(G,4)=0$, $\phi(G',4)=\frac{1}{3}$ and $\phi(G'',4)=1$.

Importantly, a rich-club with a coefficient close to 1 is a set of high degree vertices that induces a dense subgraph. In the rich-club, the membership is dictated by a ranking based on the degree of the vertices. In computing the rich-club coefficient, only the edges joining members of the rich-club are included. We introduce concepts that focus on vertices and edges that appear in many small dense subgraphs. Our aim is to identify the set of vertices that belong to many dense communities, rather than vertices with many connections. 

\subsection{Generalisation of the Rich-Club.}

Motivated by the idea of counting vertices and edges that appear in many dense subgraphs, we define two new concepts. The first concept focuses on the number of small dense subgraphs that contain a given vertex $v$. We use the notation $G'\subseteq_{i} G$ if $G'$ is an \emph{induced subgraph} of $G$, and $G\cong H$ if $G$ and $H$ are \emph{isomorphic}. In this paper, we are only interested in subgraphs $G'$ that are isomorphic to a complete graph $K_{k}$.

We formally define the \emph{vertex participation number} of a vertex $v$ in complete subgraphs of order $k$ as:
\begin{equation}
    %\xi_{(v,k)} = \sum\limits_{\substack{G'\subseteq_{i} G\\ v \in V(G')}}^{n} 1,  \mbox{ for $G'\simeq K_{k}$}.
    \xi(v,k)=|\{G'\subseteq_{i} G :v\in V(G'), G' \cong K_{k} \}|.
    \label{eq:vtxpart}
\end{equation}

Observe that, when $k=2$, the vertex participation number is the degree of the vertex. The rich-club members are the vertices with vertex participation number $\xi(v,2)> j$. Thus, the concept of vertex participation can be used to generalise the notion of a rich-club. The generalised rich-club members are the vertices  with vertex participation number $\xi(v,k)> j$. The vertices that are members of this generalised rich-club are called \emph{Super rich-club members}. The generalised rich-club is called the \emph{Super rich-club}. 

Our second concept focuses on the number of dense subgraphs that contain a given edge $e=\{u,v\}$. We focus on complete subgraphs of order $k$. We formally define the \emph{edge participation number} of a pair of vertices $\{u,v\}$ in complete subgraphs of order $k$ as:  
\begin{equation}
    %\xi_{(\{u,v\},k)} = \sum\limits_{\substack{G'\subseteq G\\ \{u,v\} \in E(G')}}^{n} 1, \mbox{ for $G'\simeq K_{k}$}.
    \xi(\{u,v\},k) = |\{G'\subseteq_{i} G : \{u,v\}\in E(G'), G'\cong K_{k} \}|.
    \label{eq:edgepart}
\end{equation}

The \emph{rich edge-club} members are the set of vertices incident to edges with high edge participation number.

\subsection{Contributions and Paper Structure.}
In Section~\ref{sec:measure}, we will discuss the two new concepts used in this paper. Note that the graphs used throughout this paper are \emph{simple} graphs. We will compare the concepts of vertex participation and edge participation with the rich-club using examples. We build upon these concepts and define the generalised rich-club coefficient, which also further extends to a weighted variant. The coefficient of the rich edge-club is also introduced. We demonstrate both concepts on a healthy human brain network. However, in real-world data collection, data is often noisy or incomplete, resulting in some missing edges in a network. To address this, we consider subgraphs that are very dense and are near complete which we call \emph{pseudo}-$K_{k}$s. 

First, in Section~\ref{sec:vertex} we look at the generalised rich-club and compare it to that of the known rich-club in brain networks. We also experiment the concept of vertex participation on randomised \emph{Erd\"{o}s R\'{e}nyi} and \emph{Watts-Strogatz} small-world networks. Results from the experiments show that the generalised rich-club concept can identify vertices that do not appear in the rich-club. This indicates that our generalised rich-club concept could identify entities that could potentially be important in real-world networks. %showing direct relationships between smaller communities? 

Then in Section~\ref{sec:edge}, we build upon the concept of edge participation and demonstrate this concept on a brain network. Results show that this concept identifies a special subset of the rich-club vertices termed \emph{SUpernodes}. A general algorithm used in identifying the set of SUpernodes is given in Section~\ref{sec:edge}, showing interesting similarities to that of the rich-club in the brain that were studied by~\citet{heuvel2011} and~\citet{verhelst2018}. 

\section{Vertex and Edge Participation}
\label{sec:measure}

A complete subgraph represents a cohesive set of vertices in a graph, and can be thought of as a most tight-knit community. Counting complete subgraphs gives a measure on the cohesiveness of a graph, or on the cliquishness of communities in a network. Hence, there has been a lot of research done on counting complete subgraphs~\cite{dixit2014,eden2018,jain2020, jain2017,rasmussen1998,yang2019}. However, little work has been done in counting vertices and edges that belong to one or more complete subgraphs. Instead of counting complete subgraphs of order $k$ in $G$, we count the number of times a vertex or an edge appear in complete subgraphs of $G$. We call these numbers the \emph{vertex participation} $\xi(v,k)$ and \emph{edge participation} $\xi(\{u,v\},k)$, $k\geq  2$. 

When $k=2$, our notion of the super rich-club coincides with the \emph{members} of the rich-club which are selected based on the ranking of the degrees of the vertices (i.e., the number of $K_{2}$s they are incident to). We will consider and discuss both the vertex and the edge participation numbers for $k>2$. First, we will develop the notion of a generalised rich-club based on the concept of vertex participation. We will then introduce the notion of a rich edge-club based on the concept of edge participation. Both concepts will be further demonstrated on the brain network to illustrate the difference between the rich club and the super rich-club. 

\subsection{Vertex Participation}

The vertex participation $\xi(v,k)$ generalise the notion of a rich-club. Membership of the Super rich-club is based on $\xi(v,k)$. The rich-club is equivalent to the Super rich-club when  $k=2$, as $\xi(v,k)$ is the degree of the vertex $v$. Thus, the Super rich-club is the set of vertices of high vertex participation, but with the edges weighted by the number of $K_{k}$s a given pair of vertices are incident to, or the \emph{edge-participation number}.

\begin{figure}[H]
    \centering
    \begin{tikzpicture}
      [scale=.6,auto=left,every node/.style={circle,fill=black}]
      \node[fill=red,label=10:$v_{1}$] (n0) at (0,0) {};
      \node[fill=red,label=135:$v_{2}$] (n1) at (-1.5,-1) {};
      \node[label=left:$v_{7}$] (n2) at (-3,-1) {};
      \node[fill=red,label=45:$v_{3}$] (n3) at (1.5,-1) {};
      \node[label=right:$v_{9}$] (n4) at (3,-1) {};
      \node[fill=red,label=225:$v_{4}$] (n5) at (-1.5,-3) {};
      \node[label=left:$v_{8}$] (n6) at (-3,-3) {};
      \node[fill=red,label=315:$v_{5}$] (n7) at (1.5,-3) {};
      \node[label=right:$v_{10}$] (n8) at (3,-3) {};
      \node[fill=red,label=190:$v_{6}$] (n9) at (0,-4) {};
      \node (n10) at (-1,1.5) {};
      \node (n11) at (0,1.5) {};
      \node (n12) at (1,1.5) {};
      \node (n13) at (-1,-5.5) {};
      \node (n14) at (0,-5.5) {};
      \node (n15) at (1,-5.5) {};
      \node[fill=none] (n16) at (0,-7) {$G$};
      
      \path[draw,line width=1mm] (n0) edge[red] (n1);
    \path[draw,line width=1mm] (n0) edge[red] (n3);
     \path[draw,line width=1mm] (n1) edge[red] (n5);
    \path[draw,line width=1mm] (n3) edge[red] (n7);
    \path[draw,line width=1mm] (n5) edge[red] (n9);
    \path[draw,line width=1mm] (n9) edge[red] (n7);
    \path[draw,line width=1mm] (n0) edge[red] (n9);
    
    \node [rotate=-50][fill=none,draw,blue, dashed,inner sep=1pt, circle,yscale=.8, fit={(n1) (n2) (n5) (n6)}] {};
    \node [rotate=50][fill=none,draw,blue, dashed,inner sep=1pt, circle,yscale=.8, fit={(n3) (n4) (n7) (n8)}] {};
      
      \foreach \from/\to in {n0/n10,n0/n11,n0/n12,n0/n3,n0/n9,n9/n13,n9/n14,n9/n15,n9/n5,n9/n7,n1/n2,n1/n6,n1/n5,n2/n6,n5/n6,n3/n4,n3/n7,n3/n8,n4/n7,n7/n8,n4/n8,n2/n5,n0/n1}
      \draw (\from) -- (\to);
      
    \end{tikzpicture}
    \caption{An example illustrating the difference between the notion of a rich-club for  and the generalised rich-club. Rich-club members are the red vertices, while the Super rich-club is circled in blue dotted lines.}
    \label{fig:super rich club}
\end{figure}
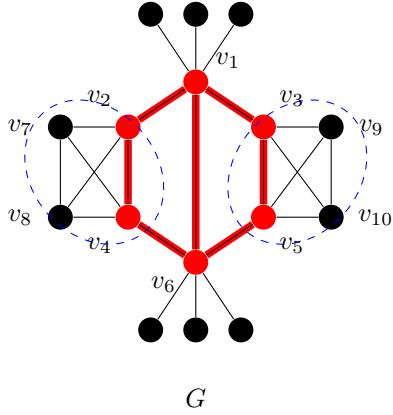

Figure~\ref{fig:super rich club} illustrates the difference between the notion of a rich-club ($j>3$) and the generalised rich-club  ($\xi(v,3)>2$). In the graph $G$, the rich-club \emph{members} (red) are $v_{1},v_{2},v_{3},v_{4},v_{5}$ and $v_{6}$, as $d(v_{1})$, $d(v_{2})$, $d(v_{3})$, $d(v_{4})$, $d(v_{5})$, $d(v_{6})>3$. In contrast, the Super rich-club are the vertices (circled in blue dotted lines) $v_{2}$, $v_{3}$, $v_{4}$, $v_{5}$, $v_{7}$, $v_{8}$, $v_{9}$ and $v_{10}$ as $\xi(v_{2},3)=\xi(v_{3},3)=\xi(v_{4},3)=\xi(v_{5},3)=\xi(v_{7},3)=\xi(v_{8},3)=\xi(v_{9},3)=\xi(v_{10},3)=3$. In this example, not all rich-club members are Super rich-club members.

The rich-club coefficient is the density of the subgraph induced by the rich-club vertices. We generalise the rich-club coefficient to give us a density measure for the generalised rich-club. We define the \emph{participation coefficient} $\mathtt{c}(G,k,j)$ as:
\begin{equation}
 \mathtt{c}(G,k,j)=\dfrac{E_{\xi(v,k)>j}}{\binom{N_{\xi(v,k)>j}}{2}}   
\end{equation}
where $E_{\xi(v,k)>j}$ and $N_{\xi(v,k)>j}$ are the number of edges and vertices in the Super rich-club respectively. When $k=2$, $\phi(G,j)=\mathtt{c}_{w}(G,2,j)$ is the rich-club coefficient $\phi(G,j)$. We also define a weighted version of this participation coefficient $\mathtt{c}_{w}(G,k,j)$ for $w>j$. This measures the proportion of the edge weights of the Super rich-club to the maximal weights in a given graph $G$. This coefficient is defined as:
\begin{equation}
    \mathtt{c}_{w}(G,k,j)=\dfrac{\sum w_{(E_{\xi(v,k)>j})}}{\sum w_{>j}}
\end{equation}
where $w_{(E_{\xi(v,k)>j})}$ are the weights of the edges of the Super rich-club and $w_{>j}$ are the edge weights $>j$ in a given graph $G$.

\subsection{Edge Participation}

In this section, we look at the number of $K_{k}$s and edge $\{u,v\}$ a part of --- we call this \emph{edge participation number} $\xi(\{u,v\},k)$. In the case of the rich-club, each edge has a participation number $\xi(\{u,v\},2)$ of 1, since each edge participates in exactly one $K_{2}$. When $k> 2$, an edge can participate in multiple $K_{k}$s. The edges that have high edge participation number $\xi(\{u,v\},k)>j$ are called \emph{rich-edge club}. The resulting edges form a weighted subgraph, where the edges are weighted by the edge participation number in $K_{k}$s.

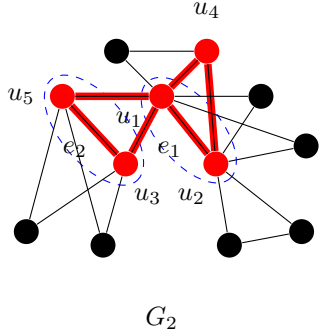
\begin{figure}[H]
    \centering
    \begin{tikzpicture}
      [scale=.6,auto=left,every node/.style={circle,fill=black}]
    \node (n0) at (0,3) {};
    \node[label=above:$u_{4}$,fill=red] (n1) at (2,3) {};
    \node[label=195:$u_{1}$,fill=red] (n2) at (1,2) {};
    \node[label=left:$u_{5}$,fill=red] (n3) at (-1.2,2) {};
    \node (n4) at (3.2,2) {};
    \node[label=280:$u_{3}$,fill=red] (n5) at (0.2,0.5) {};
    \node (n6) at (-2,-1) {};
    \node (n7) at (-0.3,-1.3) {};
    \node (n8) at (4.2, 0.9) {};
    \node[label=250:$u_{2}$,fill=red] (n9) at (2.2, 0.5) {};
    \node (n10) at (4.1,-1) {};
    \node (n11) at (2.5,-1.3) {};
    \node[fill=none] (n12) [below=of n2] {};
    \node[fill=none] (n13) [below=of n12] {$G_{2}$};
    
   \node [rotate=-50][fill=none,draw,blue, dashed,inner sep=0.5pt, circle,yscale=.5, fit={(n3) (n5)}] {};
   \node [rotate=-50][fill=none,draw,blue, dashed,inner sep=0.5pt, circle,yscale=.5, fit={(n2) (n9)}] {};

   \path (n2) edge node[below left, fill=none] {$e_{1}$} (n9);
   \path (n3) edge node[below left, fill=none] {$e_{2}$} (n5);
   \path[draw,line width=1mm] (n2) edge[red]  (n5);
   \path[draw,line width=1mm] (n2) edge[red] (n1);
   \path[draw,line width=1mm] (n2) edge[red] (n9);
   \path[draw,line width=1mm] (n9) edge[red] (n1);
   \path[draw,line width=1mm] (n2) edge[red] (n3);
   \path[draw,line width=1mm] (n3) edge[red] (n5);
   
 \foreach \from/\to in {n0/n1,n1/n2,n0/n2,n2/n3,n3/n5,n2/n5,n5/n6,n5/n7,n2/n4,n2/n8,n9/n4,n9/n8,n10/n11,n11/n9,n9/n10,n1/n9,n2/n9,n3/n6,n3/n7}
  \draw (\from) -- (\to);
    \end{tikzpicture}
    \caption{An example illustrating the difference between the notion of a rich-club for and our rich edge-club. Rich-club members are the red vertices, while the rich edge-club are circled in blue dotted lines.}
    \label{fig:rich edge club exp}
\end{figure}

Figure~\ref{fig:rich edge club exp} compares the rich club and the rich edge-club. In this example, we set $k=3$ and $j>2$. In graph $G_{2}$, the rich-club \emph{members} (red) have degree at least 3, and the rich-club is the induced subgraph shown in red. However, the edges $e_{1}=\{u_{1},u_{2}\}$ and $e_{2}=\{u_{3},u_{5}\}$ form the rich edge-club (circled in blue dotted lines) since each of these edges are incident to at least 3 $K_{3}$s. The vertices $u_{1},u_{2},u_{3}$ and $u_{5}$ are the rich edge-club members of $G_{2}$.

Similar to the weighted rich-club coefficient, we determined a measure for the rich-edge club, we call this the \emph{edge-club coefficient} $\mathtt{c}_{e}(G,k,j)$. In the rich edge-club, we measure the proportion of the sum of edge club weights to the total weights with respect to the edge participation. We compute this measure using the edge weights that are the edge participation of $K_{k}$s in $G$. This coefficient is defined as:
\begin{equation}
    \mathtt{c}_{e}(G_{w},k,j)=\dfrac{\sum\limits_{e \in E(G')} \xi(\{u,v\},k)>j}{\sum\limits_{e \in E(G)} \xi(\{u,v\},k)} = \dfrac{\sum\limits_{e \in E(G')} \xi(\{u,v\},k)>j}{\binom{k}{2}\times \#G'_{\cong K_{k}}}
\end{equation}
where $\xi(\{u,v\},k)>j$ are the edge participations $>j$ and $\#G'_{\cong K_{k}}$ is the number of subgraphs $K_{k}$ in a given graph $G$. A summary of the similarities and differences between the different rich-club concepts and coefficients are given in Table~\ref{table:summ}.

\begin{table}
\begin{center}
\resizebox{\textwidth}{!}{\begin{tabular}{|P{0.33\linewidth} | P{0.33\linewidth}|P{0.33\linewidth}|}
\hline
 \textbf{Rich-club} & \textbf{Generalised Rich-club} & \textbf{Rich Edge-Club} \\
 \hline
    Vertices are ranked by vertex degree (i.e.\ vertex participation in $K_{2}$s).  
    & Vertices are ranked by the vertex participation  in $K_{k}$s. 
    & Edges are ranked by edge participation in $K_{k}$s. \\
    \hline
    The vertices with degree $>j$ are rich-club members. 
    &  The vertices with vertex participation $>j$ are  the Super rich-club members. 
    & The edges with edge participation $>w$ are members of the rich edge-club. \\ 
    \hline
    The rich-club graph is the graph induced by the rich-club members. 
    & The generalised rich-club graph is the graph induced by the super rich-club members but with the edges assigned weights representing edge-participation. 
    & The rich edge-club graph is the subgraph that has the same vertex set as the original graph but the edges are the rich edge-club members weighted by their edge-participation. \\
    \hline
    Rich-club coefficient: 
    \begin{equation*}
    \begin{split}
    \phi(G,j)  = \dfrac{E_{>j}}{\binom{N_{>j}}{2}} \\
      = \dfrac{E_{\xi(v,2)>j}}{\binom{N_{>j}}{2}}.  
    \end{split}
    \end{equation*} 
    & Participation coefficient: \[\mathtt{c}_{(G,j)}=\dfrac{E_{\xi(v,k)>j}}{\binom{N_{\xi(v,k)>j}}{2}}.\]
    Weighted participation coefficient: \[\mathtt{c}_{w}(G,j)=\dfrac{\sum w_{(E_{\xi(v,k)>j})}}{\sum w_{>j}}.\] 
    & Edge-club coefficient: 
    \newline
    \[
    \mathtt{c}_{e(G_{w},k,j)}=\dfrac{\sum\limits_{e \in E(G')} w_{>j}}{\binom{k}{2}\times \#G'_{\cong K_{k}}} 
    \] \\
    \hline
\end{tabular}}
\caption{A comparison between the different rich-club concepts.}
\label{table:summ}
\end{center}
\end{table}

\subsection{Thresholding}

Members of the rich-club, Super rich-club and rich edge-club are selected based on a threshold. Each of the club memberships is determined based on a different ranking or selection criteria. In the rich-club, vertices are ranked by vertex degree. When thresholding on the vertex degree, we retain only vertices of degree $>j$ and disregard all vertices of lower degree. The resulting set of vertices induce a subgraph of the graph $G$. 

In the super rich-club, we rank vertices by the vertex participation number $\xi(v,k)$. We retain only vertices that have high vertex participation numbers. The threshold $j$ need not be the same for the rich-club and the vertex participation. The resulting club is a weighted subgraph where the edges are weighted by the edge participation number.

In the rich-edge club, edges are ranked by the edge participation number $\xi(\{u,v\},k)$. Each edge is weighted by the edge participation number. We retain edges which have weight greater than $w$ to create the graph $G_{w}$. This method of thresholding will result in a graph with the \emph{same} set of vertices and a subset of the edges. This new graph $G_{w}$ may be a disconnected graph containing multiple connected components, or just a single connected graph. This graph $G_{w}$ represents a community of pairs each pair being heavily affiliated in multiple mutual small clubs.  

\section{Super Rich-Club}
\label{sec:vertex}

The notion of the rich-club focuses on the degree of the vertices in a graph $G$. We generalise the notion of the rich-club and focus on vertices that belong to many complete subgraphs. One of our contributions in this paper is the introduction of the concept of vertex participation $\xi(v,k)$, for $k\geq 2$. Recall that when $k=2$, this is the degree of a vertex $v$ since $\xi(v,2)=1$.   

In real world data, some connections may be missing or incorrectly recorded due to the data being noisy or incomplete. It is possible that some subgraphs are in fact complete graphs with some missing edges and could potentially be significant. Thus, it is important to also consider subgraphs that are `almost' a complete graph or \emph{pseudo}-$K_{k}$s. In this paper, we will consider the vertex participation in pseudo-$K_{k}$s to identify the super rich-club in a given graph $G$. 

We demonstrate the concept of vertex participation on real human brain data with different resolutions (details on data collection can be found in Section~\ref{sec:supp}). We demonstrate the case for $k=5$ and compute the vertex participation number in pseudo-$K_{5}$s in $G$. Recall that any vertex $v$ in $G$ can belong to multiple $K_{5}$s at once. We denote the members of the rich-club as the set $R$ and the members of the Super rich-club as the set $S$. First, we collect the subset $R$ of vertices with degree $>j$. We then order and collect the subset $S$ of vertices with vertex participation $\xi(v,5)>j'$. We select $j$ and $j'$ such that $|R|$ and $|S|$ are as close as possible for comparison between elements.  

The concept of vertex participation is also demonstrated on randomised  \emph{Erd\"{o}s R\'{e}nyi} and \emph{Watts-Strogatz} small-world graphs. The Erd\"{o}s R\'{e}nyi graphs were generated based on the different brain graph resolutions and density. The orders of the Watts-Strogatz graphs were chosen to be similar to the brain data but with varying densities. The subset $R$ and $S$ were found in each graph and elements within each set were compared. Again, we select $j$ and $j'$ to make $|R|$ and $|S|$ as close as possible to compare the elements between each set. 

\subsection{Results}
Results from each of the experiments are presented as follows:

\begin{table}[H]
\begin{center}
\resizebox{\textwidth}{!}{\begin{tabular}{|P{0.25\linewidth}|P{0.25\linewidth} | P{0.25\linewidth}|P{0.25\linewidth}|}
\hline
  & \textbf{Brain 68} & \textbf{Brain 114} & \textbf{Brain 219} \\
 \hline
  \textbf{Rich-club $R$} & & &  \\
  \hline
    $d(v)$ & $>65$ & $>105$ & $>177$ \\
    \hline
    Number of members & 10 &  19 &  34 \\
    \hline
    \textbf{Super rich-club $S$} & & &  \\
    \hline
    $\xi(v,k)$ & $>2000000$ & $>14900000$ & $>203790000$ \\
    \hline
    Number of members &  10 &  19 &  34 \\
    \hline
    Proportion of elements in both $R$ and $S$ & 0.79 & $\approx 0.77$ & $\approx 0.94$ \\
    \hline 
    Proportion of elements in $R$ and not in $S$ & 0.21 & $\approx 0.23$ & $\approx 0.06$ \\
    \hline 
    Proportion of elements in $S$ and not in $R$ & 0.21 & $\approx 0.23$ & $\approx 0.06$ \\
    \hline 
\end{tabular}}
\caption{Result comparison between the rich-club (set $R$) and the Super rich-club  (set $S$) of different brain resolutions.}
\label{table:brain}
\end{center}
\end{table}

\begin{table}[H]
\begin{center}
\resizebox{\textwidth}{!}{\begin{tabular}{|p{0.25\linewidth}|p{0.25\linewidth} | p{0.25\linewidth}|p{0.25\linewidth}|}
\hline
  & \textbf{Random 68} & \textbf{Random 114} & \textbf{Random 219} \\
 \hline
    \textbf{Rich-club $R$} & & & \\
    \hline
    $d(v)$ & $>61$ & $>95$ & $>147$  \\
    \hline
    Number of members & 17 & 32 &  57 \\
    \hline
    \textbf{Generalised rich-club $S$} & & & \\
    \hline 
    $\xi(v,k)$ & $>970000$ & $>3979700$ & $>8840000$  \\
    \hline
    Number of members &  17 &  32 &  57 \\
    \hline
    Proportion of elements in both $S$ and $R$ & $\approx 0.91$ & $\approx 0.92$ & $\approx 0.93$  \\
    \hline 
    Proportion of elements in $S$ and not in $R$ & $\approx 0.09$ & $\approx 0.08$ & $\approx 0.07$ \\
    \hline 
    Proportion of elements in $R$ and not in $S$ & $\approx 0.09$ & $\approx 0.08$ & $\approx 0.07$ \\
    \hline 
\end{tabular}}
\caption{Result comparison between the rich-club (set $R$) and the Super rich-club  (set $S$) of generated random graphs with similar properties to the brain data.}
\label{table:random}
\end{center}
\end{table}

Table~\ref{table:brain} compares the rich-club (set $R$) and Super-rich club (set $S$) in graphs from the same brain data with three different resolutions (68, 114 and 219 respectively). The different brain resolutions is equivalent to the order of a graph. In each graph, some vertices belonged to both the rich-club and super rich-club, but there were also vertices that only belonged to the rich-club or super rich-club. This demonstrates that our generalised rich-club gives new insights into highly connected vertices. Similarly, Table~\ref{table:random} also shows the comparison in random graphs that have the same density as the brain graph for each resolution. 

\begin{table}[H]
\begin{center}
\resizebox{\textwidth}{!}{\begin{tabular}{|p{0.2\linewidth}|p{0.2\linewidth} | p{0.2\linewidth}|p{0.2\linewidth}|p{0.2\linewidth}|}
\hline
  & \textbf{$\delta(G)$=0.25} & \textbf{$\delta(G)$=0.50} & \textbf{$\delta(G)$=0.75} & \textbf{$\delta(G)$=0.90} \\
  \hhline{|=|=|=|=|=|}
  \textbf{$|G|$=50, $N$=10} &  &  & & \\
  \hhline{|=|=|=|=|=|}
  mean ($\mu$) of swap distance & 45.5 & 41.8 & 35.2 & 31.6 \\
  \hline
  standard deviation ($\sigma$) of swap distance & 1.509 & 1.989 & 1.989 & 2.119  \\
  \hhline{|=|=|=|=|=|}
   \textbf{$|G|$=100, $N$=10} & &  &  & \\
  \hhline{|=|=|=|=|=|}
  mean ($\mu$) of swap distance & 93.5 & 88.3 & 81.3 & 81 \\
  \hline
  standard deviation ($\sigma$) of swap distance & 2.068 & 2.584 & 4.620 & 2.404 \\
  \hhline{|=|=|=|=|=|}
   \textbf{$|G|$=200, $N$=10} &  &  &  & \\
  \hhline{|=|=|=|=|=|}
  mean ($\mu$) of swap distance & 192.8 & 184.6 & 176.6 & 185.5 \\
  \hline
   standard deviation ($\sigma$) of swap distance & 2.201 & 2.459 & 3.406 & 4.673 \\
   \hline 
\end{tabular}}
\caption{Result comparison between the rich-club (set $R$) and the Super rich-club (set $S$) in generated Watts-Strogatz small-world random networks using edit distance between elements (members) of each set.}
\label{table:random watts}
\end{center}
\end{table}

To further test the concept of vertex participation, we generated the \emph{Watts-Strogatz} small-world random graphs of order $|G|=50,100$ and 200 with different densities $\delta(G)=0.25$, 0.5, 0.75 and 0.9. We generated a sample of $N=10$ for each order with different densities and compared both sets $R$ and $S$. The ordering of the vertices differ when ranked by degree (for $R$) and by the vertex participation (for $S$). Thus, we compared the ordering of the vertices using the swap distance between the different rankings. The mean $\mu$ and standard deviation $\sigma$ of the swap distance was computed and presented in Table~\ref{table:random watts}. 

\subsection{Discussion}
We have demonstrated the difference between both the rich-club and Super rich-club based on the human brain data of different resolutions ($n=68,114,219$). We ordered the degree of the vertices $d(v)$ for each resolution and selected the rich-club members based on degree $>j$. The results show that not all high degree vertices belong in many cliques. This could potentially reveal vertices (regions) that play important local roles in the brain. 

\begin{figure}[H]
    \centering
    \subfloat[Proportion of elements in $S$ \\ ~~~~~~that are also in $R$.]{\includegraphics[width=0.5\textwidth]{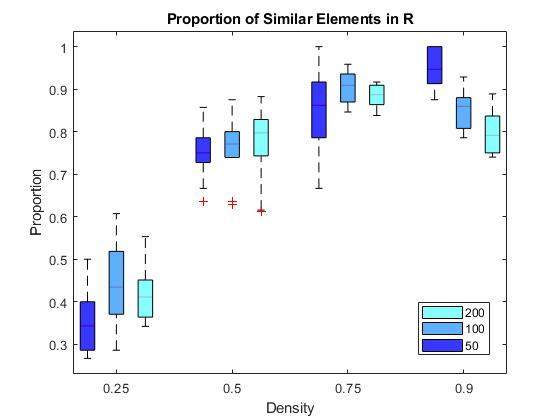}\label{fig:f1}}
    \hfill
    \subfloat[Proportion of elements in $R$ \\~~~~~~ that are also in $S$.]{\includegraphics[width=0.5\textwidth]{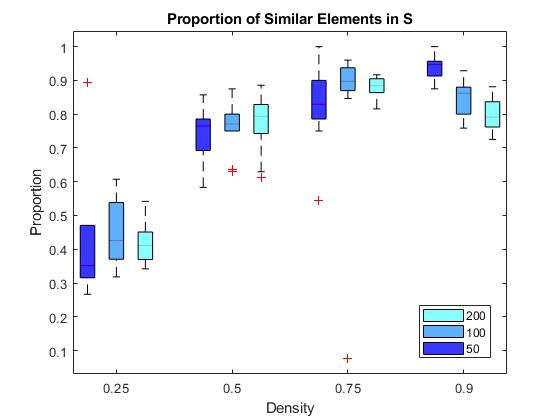}\label{fig:f2}}
    \caption{Boxplots showing the proportion of common elements between $S$ and $R$.}
    \label{fig:boxplots}
\end{figure}

The results from Table~\ref{table:brain} show the similarity and differences in elements between both sets. Notice that no two sets from either resolution contain the exact same elements. In each resolution, there are at least $\approx 6\%$ of the elements that are not the same. This suggests that not all vertices of high degree belong in the set $S$, similarly not all vertices with high vertex participation belong in the set $R$. The result was also consistent when demonstrated on the Erd\"{o}s R\'{e}nyi random graphs generated with the same density as each of the different brain resolution. The results from the random graphs are also shown in Table~\ref{table:random}. 

\begin{figure}[H]
    \centering
    \includegraphics[width=0.9\textwidth]{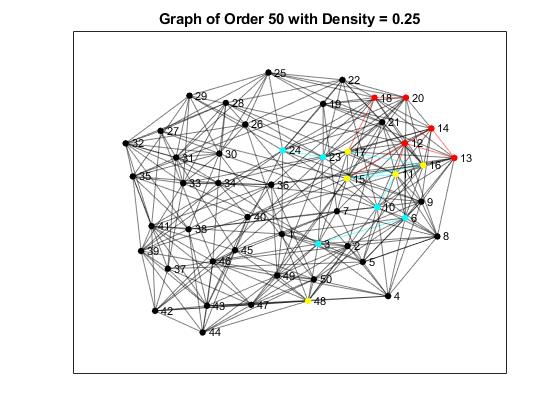}
    \caption{Example of a Watts-Strogatz small world graph $G$ of order 50 and $\delta(G)=0.25$. The set $R$ is shown in red, the set $S$ is shown blue and the common elements are shown in yellow.}
    \label{fig:exampleRScommon}
\end{figure}

The concept of vertex participation was further demonstrated on the \emph{Watts-Strogatz} small world graphs. The results that follow were consistent with the results from Table~\ref{table:random} and Table~\ref{table:random watts}. Figure~\ref{fig:boxplots} shows the proportion of common elements (members) between the rich-club and Super rich-club. An interesting observation that emerged from Figure~\ref{fig:boxplots} is that the greatest difference between elements of the rich-club and Super rich-club occurs in low density graphs. An example illustrating this scenario is shown in Figure~\ref{fig:exampleRScommon}.

The results also show that even at a high density of $\delta(G)=0.9$, the proportion of common elements between rich-club and the Super rich-club falls below 1. Further, at a low density of $\delta(G)=0.25$, the mean swap distance was greater compared to the higher density graphs. The results also suggest that the concept of vertex-participation was able to identify vertices that may have local importance in a network.  

\section{Rich Edge-Club}
\label{sec:edge}

We investigate edge participation in brain networks and show the thresholding technique on the edge participation number $\xi(\{u,v\},k)$. This thresholding is based on the edge ordering by edge participation, where we only consider the subset of edges that contain edge weights $>w$. Recall that a \emph{pseudo}-$K_{k}$ is an almost complete graph with some edges missing. We introduce a weighted pseudo-$K_{k}$ such that each edge is assigned a numeric value. Due to the possible presence of noise in real life data collection which may result in missing edges in the data, we will consider the edge participation of pseudo-$K_{k}$s in $G$.

First, we present a general idea of the algorithm used in counting pseudo-$K_{k}$s including exact $K_{k}$s for any given graph $G$. The general idea is as follows: Suppose we are given a weighted graph $G=(V,E)$, we first compute the edge participation numbers in small dense subgraphs of $G$. We then create a new graph $G_{i}$ with the same set of vertices of $G$ but each edge is weighted by the number of subgraphs the edge participates in. Edges with weight $<w$ are discarded ($w$ changes with each iteration). This process is applied iteratively until the resulting graphs consists of isolated vertices and highly weighted (connected) component(s). The non-isolated vertices are the set of rich edge-club vertices.

\subsection{MRI Data and Connectivity matrices}

Graphs were obtained from human brain data as follows. We utilised structural connectivity data derived from diffusion \textsc{mri} provided by the Human Connectome Project (\textsc{hcp}; \url{http://www.humanconnectome.org}) from the Washington University-University of Minnesota (\textsc{wuminn}) consortium, including 484 healthy participants from the Q4 release. The pre-processing of this data is described by~\citet{lrhs=2019,phpdc-2020}. Structural connectivity (\textsc{sc}) adjacency matrix edge weights were defined as the number of reconstructed streamlines between two regions of interest derived from diffusion \textsc{mri} deterministic tractography~\cite{lrhs=2019}. This measurement reflects the number of tracts connecting two brain regions, and was chosen for maximum comparability with existing diffusion \textsc{mri} studies~\cite{jeurissen19}. \textsc{sc} connectomes were constructed for each individual subject and used for subsequent analyses. A group-average weighted \textsc{sc} matrix was then computed using the sum of all matrix values across individuals divided by 484. All post-processing of adjacency matrices was completed using Matlab 2018a (\url{http://www.mathworks.com}) and Python (version 3.0). A detailed description of \textsc{mri} data acquisition parameters can be found in supplementary material.

\subsubsection{Parcellation}

The nodes (vertices) of the brain graph are determined by the brain parcellation (atlas). Connectivity matrices were parcellated using a multi-resolution (68, 114, 219) atlas derived from the Desikan-Killiany cortical atlas (excluding sub-cortical areas) as proposed by~\citet{cgmt-2012}. Each iterative analysis was repeated at all parcellation resolutions to ensure that effects were independent of spatial scale. 

\subsection{Algorithm}

\begin{figure}[H]
\centering 
\begin{tikzpicture} 
  [scale=.6,auto=left,every node/.style={circle,fill=black}]
  \node (n0) at (0,3) {};
  \node (n1) at (2,1.5) {};
  \node (n2) at (1.2,-1) {};
  \node (n3) at (-1.2,-1) {};
  \node (n4) at (-2,1.5) {};
  
  \foreach \from/\to in {n0/n1,n0/n2,n0/n3,n0/n4,n1/n2,n1/n3,n1/n4,n2/n3,n2/n4,n3/n4}
  \draw (\from) -- (\to);

   \begin{scope}[xshift=6cm]
    \node (n0) at (0,3) {};
  \node (n1) at (2,1.5) {};
  \node (n2) at (1.2,-1) {};
  \node (n3) at (-1.2,-1) {};
  \node (n4) at (-2,1.5) {};
   
   \draw[dotted] (n1) -- (n4);
   \draw[dotted] (n0) -- (n4);
   
 \foreach \from/\to in {n0/n1,n0/n3,n0/n2,n1/n2,n1/n3,n2/n3,n2/n4,n3/n4}
  \draw (\from) -- (\to);
   \end{scope}
   
   \begin{scope}[xshift=12cm]
   \node (n0) at (0,3) {};
  \node (n1) at (2,1.5) {};
  \node (n2) at (1.2,-1) {};
  \node (n3) at (-1.2,-1) {};
  \node (n4) at (-2,1.5) {};
  
  \draw[dotted] (n0) -- (n4);
  \draw[dotted] (n2) -- (n3);
  \draw[dotted] (n2) -- (n4);
  \draw[dotted] (n0) -- (n3);
   
   \foreach \from/\to in {n0/n1,n0/n2,n1/n2,n1/n3,n1/n4,n3/n4}
   \draw (\from) -- (\to);
   \end{scope}

\end{tikzpicture}
\caption{[Left to right]: Exact $K_{5}$ and pseudo-$K_{5}$-s which have median weight $w>1$, assuming each edge has $w=1$, while dotted lines represent non-edges with edge weight $w=0$.}
\label{fig:K5}
\end{figure}
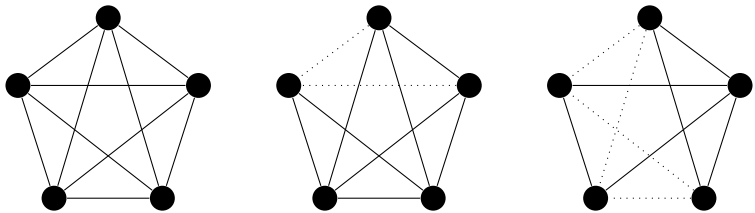

The algorithm used to identifying the set of SUpernodes in the brain data for the case $k=5$ is as follows. We first count the number of pseudo-$K_{5}$s including the number of $K_{5}$s in a given graph $G$ which we denoted $G_{0}$. These counts give a \emph{pseudo-}edge participation number for each edge in $G_{0}$. The edges are ordered by the \emph{edge participation} $\xi(\{u,v\},k)$ with respect to the current graph. An example of a pseudo-$K_{5}$ is given in Figure~\ref{fig:K5}.

\begin{figure}
\centering 
\begin{tikzpicture} 
  [scale=.6]
\begin{scope}[yshift=1cm]
 \matrix [matrix of math nodes,nodes={fill=none},fill=none, 
    left delimiter=(, 
    right delimiter=),
    ](A){ 
0 & 0 & 120  & 1000 & 300   \\
0 & 0 & 400  & 0 & 0 \\  
120 & 400 & 0  & 0 & 200 \\
1000 & 0 & 0 & 0 & 225 \\
300 & 0 & 200  & 225 & 0 \\
};
\node[above=5pt of A-1-1] (top-1) {a};
\node[above=5pt of A-1-2] (top-2) {b};
\node[above=5pt of A-1-3] (top-3) {c};
\node[above=5pt of A-1-4] (top-4) {d};
\node[above=5pt of A-1-5] (top-5) {e};

\node[left=22.5pt of A-1-1] (left-1) {a};
\node[left=22.5pt of A-2-1] (left-2) {b};
\node[left=18pt of A-3-1] (left-3) {c};
\node[left=15pt of A-4-1] (left-4) {d};
\node[left=18pt of A-5-1] (left-5) {e};

\end{scope}

\begin{scope}[xshift=10cm]
  \node[circle,fill=black,label=above:a] (n0) at (0,3)  {};
  \node[circle,fill=black, label=right:b] (n1) at (2,1.5)  {};
  \node[circle,fill=black, label=right:c] (n2) at (1.2,-1)  {};
  \node[circle,fill=black, label=left:d] (n3) at (-1.2,-1)  {};
  \node[circle,fill=black, label=left:e] (n4) at (-2,1.5) {};
  
  \draw[dotted](n2) -- (n3);
  %\draw[dotted] (n2) -- node[below,fill=none] {0} (n3); 
  \draw[dotted] (n1) --  (n4);
  \draw[dotted] (n1) --  (n3);
  \draw[dotted] (n0) --  (n1);
  
  \foreach \from/\to in {n0/n2,n0/n3,n0/n4,n1/n2,n2/n4,n3/n4}
  \draw (\from) -- (\to);
  \end{scope}
  \end{tikzpicture}
 
\begin{tikzpicture}
  [scale=0.6]
\begin{scope}[yshift = -3cm]
 \matrix [matrix of math nodes,nodes={fill=none},fill=none, 
    left delimiter=(, 
    right delimiter=),
    ](A){ 
0   & 5     & 0.5   & 100    & 0  \\
5   & 0     & 70    & 6.99   & 0 \\  
0.5 & 70    & 0     & 500    & 111.22 \\
100 & 6.99   & 500  & 0     & 98 \\
0   & 0     & 111.22  & 98     & 0 \\
};

\node[above=5pt of A-1-1] (top-1) {u};
\node[above=5pt of A-1-2] (top-2) {v};
\node[above=5pt of A-1-3] (top-3) {x};
\node[above=5pt of A-1-4] (top-4) {y};
\node[above=5pt of A-1-5] (top-5) {z};

\node[left=20.5pt of A-1-1] (left-1) {u};
\node[left=20.5pt of A-2-1] (left-2) {v};
\node[left=16pt of A-3-1] (left-3) {x};
\node[left=15pt of A-4-1] (left-4) {y};
\node[left=20.5pt of A-5-1] (left-5) {z};
\end{scope}

   \begin{scope}[xshift=10cm,yshift=-4cm]
   \node[circle,fill=black,label=above:u] (n0) at (0,3)  {};
  \node[circle,fill=black, label=right:v] (n1) at (2,1.5)  {};
  \node[circle,fill=black, label=right:x] (n2) at (1.2,-1)  {};
  \node[circle,fill=black, label=left:y] (n3) at (-1.2,-1)  {};
  \node[circle,fill=black, label=left:z] (n4) at (-2,1.5) {};
   
   \draw[dotted] (n1) -- (n4);
   \draw[dotted] (n0) -- (n4);
   
    \foreach \from/\to in {n0/n1,n0/n3,n0/n2,n1/n2,n1/n3,n2/n3,n2/n4,n3/n4}
  \draw (\from) -- (\to);
   \end{scope}
 
\end{tikzpicture}
\caption{[Top to bottom] Fix median edge weight to be $w=200$. An example illustrating a pseudo-$K_{5}$ with median edge weight $w \geq 200$ and a graph with median edge weight $w < 200$ which does not meet the threshold requirements to be considered as a pseudo-$K_{5}$.}
\label{fig:pseudo median}
\end{figure}
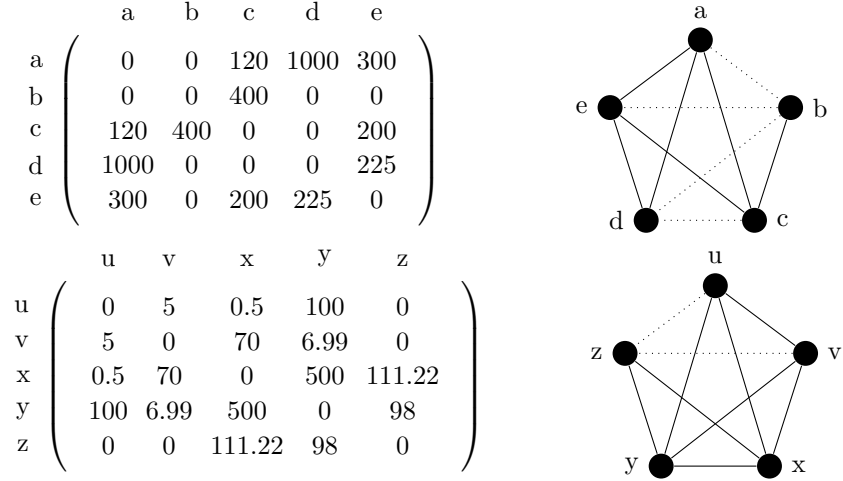

We consider all possible 5-vertex subsets in the graph. For a 5-vertex subgraph to be a pseudo-$K_{5}$, it must be connected and the \emph{median} edge weight must be greater than the threshold $w_{i}$. A non-edge is considered as an edge with weight $w=0$. In Figure~\ref{fig:pseudo median}, the top figure is an example of a weighted pseudo-$K_{5}$. A weight $w_{i}$ is assigned as a threshold at each iteration. This weight is selected such that the top $P$ percent of edges in the order of the edge participation in $G_{i}$ are retained. The sequence of percentile $P$ is illustrated in Table~\ref{table:percentile}.

At each iteration, a new graph $G_{i+1}$ is created using the edge participation of the current graph $G_{i}$ as edge weights. This process is repeated until convergence, and we are left with some highly connected weighted component(s) and isolated vertices.  The non-isolated set of vertices are the rich edge-club vertices.

\begin{table}[H]
\begin{center}
\begin{tabular}{|c|c|c|c|}
\hline
 Graph (Old) & Iteration $i$ & Percentile $P$ & Graph (New) $G_{i}$  \\
 \hline
    $G_{0}$ (Original data) & 0 & 50 (median) & $G_{1}$ \\
    $G_{1}$ & 1 & 50 (median) & $G_{2}$ \\
    $G_{2}$ & 2 & 75  & $G_{3}$ \\
    $G_{3}$ &3 & 87.5 & $G_{4}$ \\
    $G_{4}$ &4 & 93.75 & $G_{5}$ \\
    $G_{5}$ &5 & 96.875 & $G_{6}$ \\
    $\vdots$ & $\vdots$ & $\vdots$ & $\vdots$ \\
    $G_{9}$ & 9 & 99.8046875 & $G_{10}$ \\
    \hline
\end{tabular}
\caption{Table illustrating list $P$ of percentiles and the process of obtaining the new graphs $G_{i}$.}
\label{table:percentile}
\end{center}
\end{table}

Notice that iterations $i=0$ and $1$ both use the 50th percentile (median). The initial iteration works with the original graph, where the edge weights represent the number of streamlines between each pair of nodes in the network. The subsequent graphs formed are graphs with edge weights that are the edge participation number $\xi(\{u,v\},k)$. Thus, the first iteration is the graph $G_{1}$. The series of iterations comes to a halt when the algorithm returns the same number of pseudo-$K_{5}$s as the previous iteration. Each iteration eliminates `weaker' connections that do not meet the percentile threshold. Thus, the presence of a pseudo-$K_{5}$ in $G_{i}$ does not necessarily mean it is also a pseudo-$K_{5}$ in $G_{i+1}$.

\subsection{Results and Discussion}
\label{edge results}

\begin{figure}
\begin{center}
\includegraphics[width=1\textwidth]{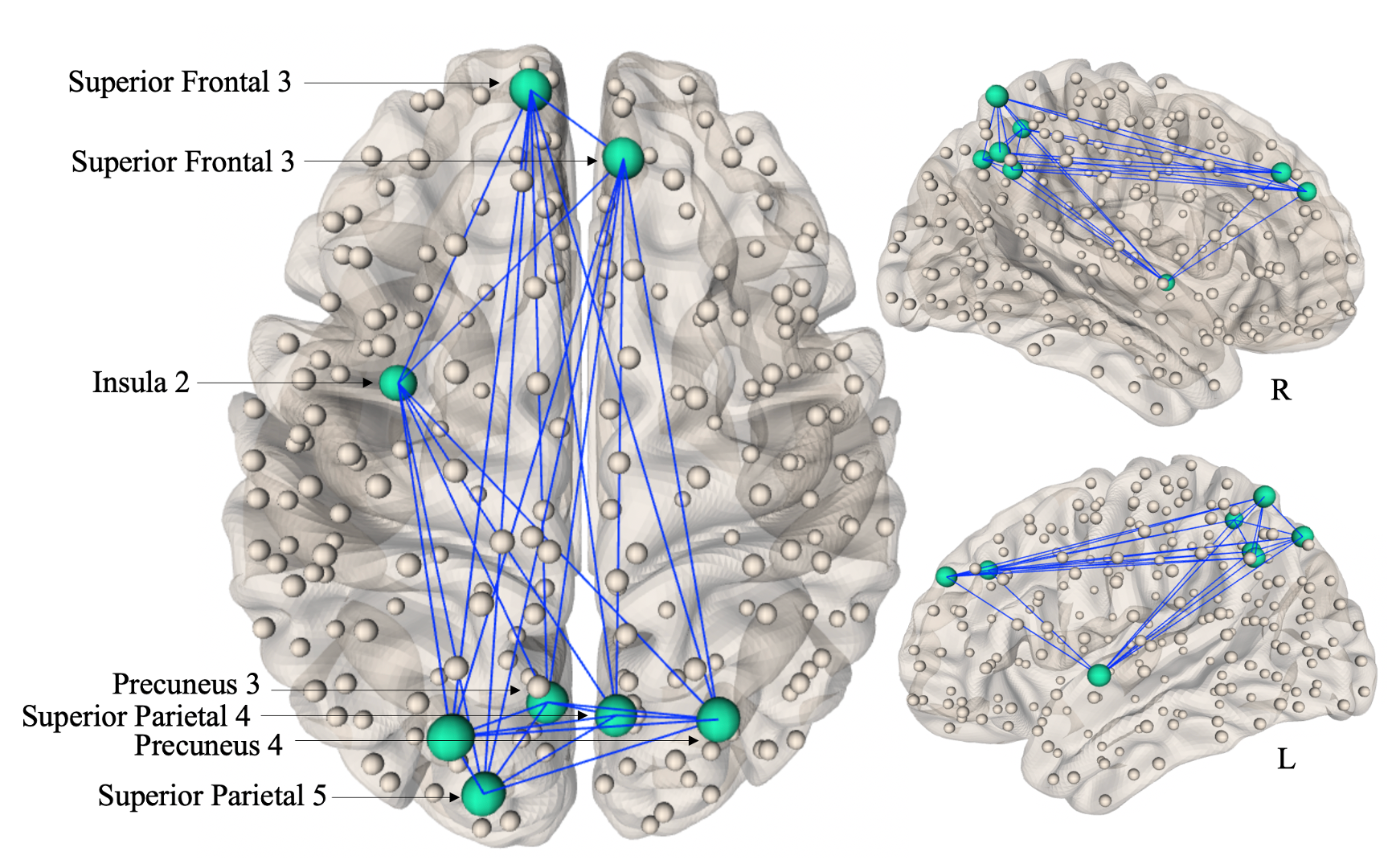}
\caption{A glass-brain plot of the spatial location of SUpernodes at a 219 resolution as identified using our concept of edge participation. This set of SUpernodes comprises bilateral superior frontal nodes 3, left insula node 2, left precuneus node 3, right superior parietal node 4, left precuneus node 4 and right superior parietal node 5.}
\label{fig:result1}
\end{center}
\end{figure}

The set of SUpernodes identified using our algorithm is given in Figure~\ref{fig:result1}. Thresholded edge participation yielded a set of eight SUpernodes from the original 219 nodes. Seven of the eight SUpernodes have been previously identified as the members of the rich-club, but by using the Desikan-Killiany atlas at a different resolution~\cite{heuvel2011}. The remaining SUpernode is the left insula node that was not previously identified as a \emph{member} of the known rich-club of the brain by~\citet{heuvel2011}. This demonstrates that the rich-club and SUpernodes are not necessarily the same set of nodes. The set of SUpernodes also include the bilateral superior frontal nodes, left precuneus nodes and the right superior parietal nodes. 

\begin{figure}[H]
\centering
\includegraphics[width=1\textwidth]{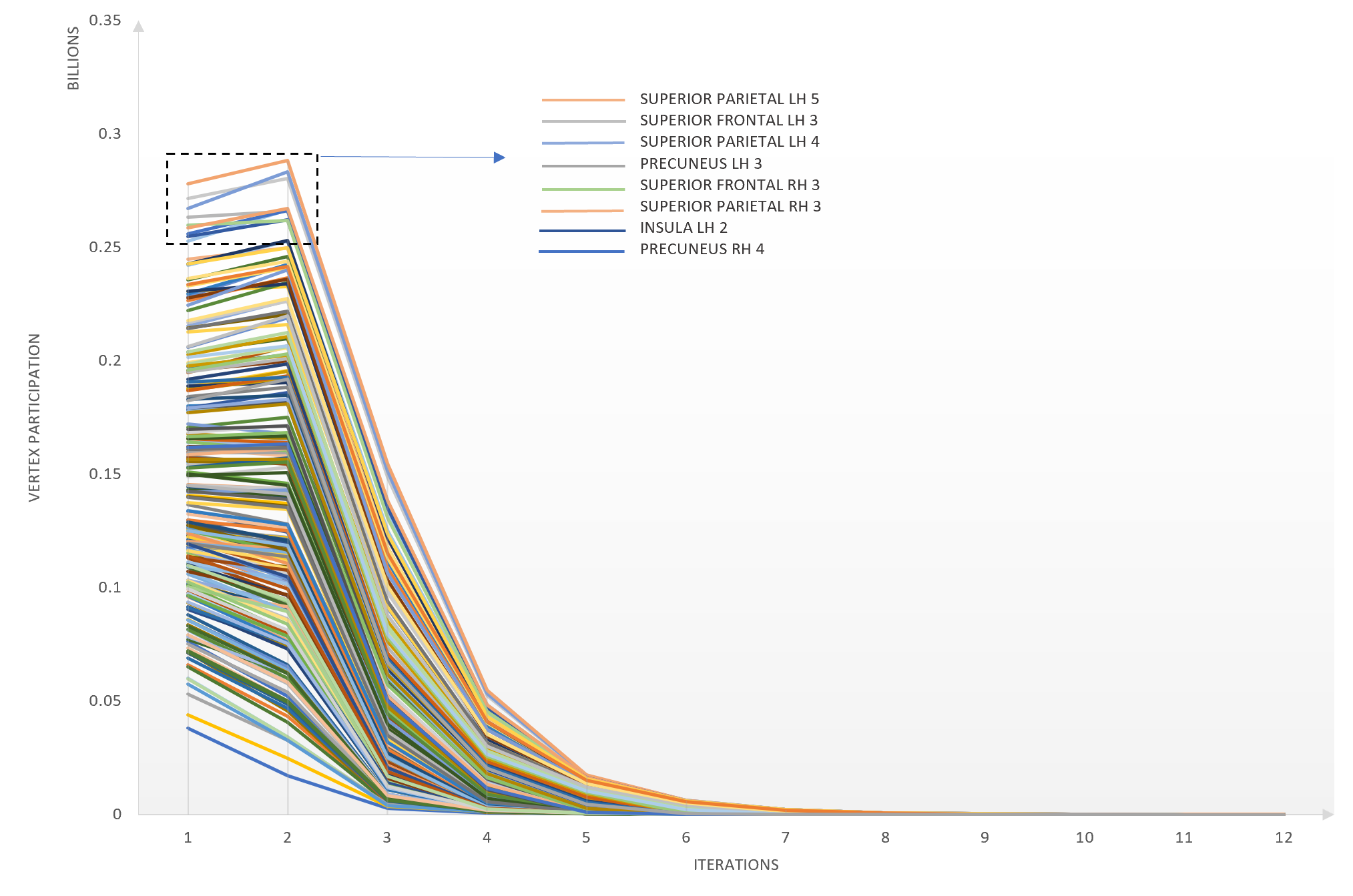}
\caption{Graph showing the reduction of the vertex participation $\xi(v,5)$ of each vertex $v$ in pseudo-$K_{5}$s after each iteration.}
\label{fig:result2}
\end{figure}

Figure~\ref{fig:result2} shows the reduction in the number of non-isolated nodes at each iteration using the thresholding technique on the concept of edge participation. Notice that the vertex participation reduces with each iteration as edges with low edge weights are discarded. The vertices that are connected to the remaining edges are members of the SUpernodes in the graph with 219 nodes.

Previously, the rich-club as identified in healthy individuals using the Desikan-Killiany atlas has included the bilateral superior frontal, putamen, thalamus, precuneus and superior parietal nodes~\cite{heuvel2011}. Our findings align with this group of nodes, albeit with the  addition of the left insula (at 219 resolution). This vertex has degree $d(v)=197$ and vertex participation $\xi(v,5)>255000000$, but was not identified using vertex degree, or largest edge-weight. This finding opens to the possibility for our novel measurement to characterise the importance of specific nodes above and beyond node degree, or edge-weight alone.

As can be seen, the concept of edge participation has was able to identify regions in the brain that could not be identified using the rich-club concept. Using this concept, we were able to show direct relationships between pairs of regions that appear in smaller communities together. This in fact could potentially be adapted to identify nodes (regions) that play an important role in certain networks.

\section{Conclusions and Future Work}
\label{sec:conclu}

In this paper, we introduced two new concepts, \emph{vertex participation} and \emph{edge participation}. These relate to subgraph counting and the rich-club concept. Existing literature has shown that the identification of the rich-club reveals important network information, for example in the brain network and also in social networks. 

The \emph{vertex participation number} $\xi(v,k)$ focused on the frequency of which a vertex $v$ appears in a subgraph $H$ of order $k$. The vertex participation is a generalisation of the concept of the \emph{rich-club}. We introduced the concept of a \emph{Super rich-club}, which is the graph induced by the vertices with high vertex participation number. 

We demonstrated the concept of vertex participation on the human brain data, randomised Erd\"{o}s-R\'{e}nyi and Watts-Strogatz small world networks. Our results have shown that the Super rich-club may contain vertices not included in the rich-club, and focuses on direct relationships between communities, rather than relationships between two people.

The \emph{edge participation number} $\xi(\{u,v\},k)$ is the frequency of which an edge $\{u,v\}$ belongs to dense subgraphs $H$ of order $k$. The \emph{rich edge-club} is the resulting set of edges and vertices. Edges in the rich edge-club represent pairs of vertices that participate in many dense subgraphs (clubs). Our algorithm used to identify the set of Supernodes in the brain utilises the edge participation.

Through this process, we identified the set of \emph{SUpernodes} in a brain network. The rich-club was identified by~\citet{heuvel2011}. Our results on brain data with a parcellation scheme of 219 regions, identified 8 SUpernodes one of which was not a member of the rich-club. This node although not the highest degree, participates in many pseudo-$K_{5}$s. We note that~\citet{heuvel2011} used a different resolution with 82 regions of interest (vertices), while our data which comprises of 219 regions.

We generalised the rich-club coefficient for the Super rich-club and also a coefficient for the edge participation. The \emph{vertex participation coefficient} measures the \emph{density} of the Super rich-club. A weighted version was defined to measure the proportion of Super rich-club weights to the maximal weights of the graph. The \emph{rich-edge club coefficient} measures the proportion of the highly ranked weights to the total weights in the graph. 

Future work will include the investigation of the generalised rich-club in social networks. Another area of future work is to investigate communities in large complex networks using the vertex- and the edge participation.

\section{Supplementary information}
\label{sec:supp}
\subsection{Sample} 

The dataset was provided by the Human Connectome Project (\textsc{hcp}; \url{http://www.humanconnectome.org})
from the Washington University-University of Minnesota (\textsc{wuminn}) consortium, including 484 healthy participants from the Q4 (500 subject) release (272 females, 212 males; age = 29.15±3.47).

\subsection{MRI data}
\textsc{mri} data acquisition were acquired using a modified 3T Siemens Skyra scanner with a 32-channel head coil.  T1- weighted structural images (\textsc{mprage}) were acquired using the following parameters:  TR = 2,400 ms, TE = 2.14 ms, flip angle = 8 ,voxel size = 0.7 mm isotropic, FOV = 224×224 mm$^{2}$ and 320 slices.  Diffusion-weighted images (\textsc{dwi}) were acquired with 270 gradient directions (multi-shell) with b-values 1000, 2000, 3000  s/mm$^{2}$.  TR = 5520 ms, TE = 89.5 ms, flip angle = 78, FOV = 210×180 mm$^{2}$, 111 slices, and voxel size = 1.25 mm isotropic.

\paragraph{Acknowledgements}
I would like to thank my colleagues and supervisors for supporting this project. I would also like to that Professor Graham Farr for his feedback on this paper. \\

\noindent
Julien Ugon’s research was supported by ARC discovery Project DP180100602.

\bibliographystyle{plainnat}
\bibliography{ref}

\end{document}